\documentclass[runningheads,a4paper]{llncs}
\usepackage[T1]{fontenc}
\usepackage{graphicx}
\usepackage{amsmath}
\usepackage{comment}
\usepackage[colorlinks=true,urlcolor=blue,citecolor=.,linkcolor=.]{hyperref}
\usepackage{csquotes}
\usepackage{listings}
\usepackage[noabbrev]{cleveref}
\usepackage{microtype}
\usepackage{hyphenat}
\usepackage{tabulary}
\usepackage{array}
\usepackage{xspace}
\usepackage{makecell}
\usepackage[american]{babel}
\usepackage{orcidlink}
\usepackage{tikz}

\usepackage{enumitem}
\newlist{rqs}{enumerate}{2}
\setlist[rqs,1]{label=RQ\arabic*.,ref=RQ\arabic*}
\setlist[rqs,2]{label=(\alph*),ref=\thequestionsi(\alph*)}
\usepackage[acronym]{glossaries-extra}
\newabbreviation{rag}{RAG}{Retrieval Augmented Generation}
\newabbreviation{llm}{LLM}{Large Language Model}
\newabbreviation{soc}{SOC}{Service-Oriented Computing}
\newabbreviation{uddi}{UDDI}{Universal Description, Discovery, and Integration}
\newabbreviation{ise}{ISE}{Information System Engineering}
\newabbreviation{gics}{GICS}{Global Industry Classification Standard}

\def \thetool {Compositio Prompto\xspace}
\begin{document}

\title{Adopting Large Language Models to Automated System Integration}
\author{
Robin D. Pesl\,\orcidlink{0000-0002-5980-9395}
}
\authorrunning{R.D. Pesl}
\institute{University of Stuttgart, Stuttgart, Germany\\
\email{robin.pesl@iaas.uni-stuttgart.de}}
\maketitle

\begin{abstract}
Modern enterprise computing systems integrate numerous subsystems to resolve a common task by yielding emergent behavior.
A widespread approach is using services implemented with Web technologies like REST or OpenAPI, which offer an interaction mechanism and service documentation standard, respectively.
Each service represents a specific business functionality, allowing encapsulation and easier maintenance.
Despite the reduced maintenance costs on an individual service level, increased integration complexity arises.
Consequently, automated service composition approaches have arisen to mitigate this issue.
Nevertheless, these approaches have not achieved high acceptance in practice due to their reliance on complex formal modeling.
Within this Ph.D. thesis, we analyze the application of \glsxtrfullpl{llm} to automatically integrate the services based on a natural language input.
The result is a reusable service composition, e.g., as program code.
While not always generating entirely correct results, the result can still be helpful by providing integration engineers with a close approximation of a suitable solution, which requires little effort to become operational.
Our research involves (i) introducing a software architecture for automated service composition using \glsxtrshortpl{llm}, (ii) analyzing \glsxtrfull{rag} for service discovery, (iii) proposing a novel natural language query-based benchmark for service discovery, and (iv) extending the benchmark to complete service composition scenarios.
We have presented our software architecture as Compositio Prompto, the analysis of \glsxtrshort{rag} for service discovery, and submitted a proposal for the service discovery benchmark.
Open topics are primarily the extension of the service discovery benchmark to service composition scenarios and the improvements of the service composition generation, e.g., using fine-tuning or \glsxtrshort{llm} agents.

\keywords{Service composition \and Service discovery \and Large language models \and OpenAPI}
\end{abstract}

\section{Introduction}

Automated service composition describes the emergence of combining multiple services to a composite service~\cite{lemos2015web}.
Automating this process yields the advantages of reduced manual effort, faster time-to-market, and agile adoption to changed business needs, resulting in an overall strategic benefit for the company.
An example would be an automotive vendor wanting to integrate roadside services like parking spot booking.
An automated approach allows for the integration of services that are not available during design time without further manual development effort.

Previous approaches to automated service composition rely on formal models, always producing correct results while requiring extensive manual modeling.
With the advent of \glspl{llm}, it has become feasible to process natural language queries and semi-structured documentation automatically, i.e., formal and natural language parts.
Employing \glspl{llm} for automated service composition could mitigate the issue of complex formal modeling by allowing developers to express their requirements in natural language while generating a code recommendation fully automatically.

This leads to our overarching research question:

\smallskip%
\noindent\fbox{\parbox{\dimeval{\linewidth-2\fboxsep-2\fboxrule}}{\centering%
    How well can \glspl{llm} be employed for automated service composition?%
}}%
\smallskip

The remainder of the paper is structured as follows.
In \Cref{sec:stoa}, we give an overview of the current literature regarding service composition in \gls{ise} and \gls{soc} and the application of \glspl{llm} for service compositions.
Then, we state our research methodology in \Cref{sec:research_methodology}.
In \Cref{sec:contributions}, we clarify our contributions.
\Cref{sec:preliminary_results} shows what we already achieved.
We elaborate on our planned work in \Cref{sec:future_work} and conclude with \Cref{sec:conclusion}.

\section{State of the Art} \label{sec:stoa}
We provide a short literature overview to motivate the topic's relevancy and explain the current state of the art.
This includes a brief description of classical service composition approaches, the subfield of service discovery, its relevance for \gls{ise}, and initial ideas to apply \glspl{llm} in \gls{soc}.

\subsection{Service Composition}
Automated service composition has been a field of research in \gls{ise} and \gls{soc} for more than two decades.
While \gls{ise} mainly focuses on its positioning in automating parts of the requirement engineering process to reduce workload and decrease time-to-market~\cite{bianchini2006ontology,dourdas2006discovering,zachos2007discovering}, \gls{soc} concentrates on its technical implementation~\cite{lemos2015web}.
This contains aspects like \textit{component access}, \textit{conversation management}, \textit{control flow}, \textit{dataflow}, and \textit{data transformation}~\cite{lemos2015web}.
While there was initially high creativity in creating solution approaches, the research slowed down.
Nevertheless, there is still a lack of a comprehensive, viable solution.

Famous classical approaches rely on AI planning, which computes a plan, i.e., a sequence, of service invocations based on formal modeling of the service and the composition requirements.
These approaches can be domain-specific~\cite{shesh03} or domain-independent~\cite{mcDermott02}.
Further approaches rely on finite state automata to model the service interaction known as the \enquote{Roman model}~\cite{berardi03,degiacomo13}.
While always producing correct results, these classical approaches require laborious and erroneous formal modeling, leading to brittle solutions and low application in practice.

In contrast, services are often documented using a semi-structured OpenAPI specification~\cite{openapi2021spec} in JSON or YAML.
It consists of general information about the service, like name or host, and the endpoints, i.e., the APIs that offer the actual functionality.
The endpoint specification again contains natural language elements like a description and structured elements like input and output schemas.
Our approach relies on OpenAPI specifications for implementations as these are the state of practice.

A subfield of automated service composition is service discovery, which identifies relevant services within a potentially vast set of all available services.
Initial ideas concentrate on centralized registries across vendors implement, e.g., in the \gls{uddi} specification~\cite{curbera2002unraveling}.
These share the same drawbacks of requiring extensive laborious manual modeling and opposing workload reduction efforts.

More recent approaches try to leverage already present OpenAPI documentation~\cite{10.1007/978-3-031-57853-3_3}.
In our work, we analyze the application of \gls{rag} with OpenAPI to realize service discovery to allow automated natural language processing, sidestepping any manual effort.

The relevancy of service composition and service discovery is backed by a long list of literature in the \gls{ise} community, e.g.,~\cite{apostolakis2023simple,bensassi2018service,dourdas2006discovering,jerbin2023request,mehandjiev2012cooperative,oriol2018ontology,wang2012formal,zachos2007discovering}.
It enables steering system design, streamlining development, dynamic system changes, reduced manual labor, increased scalability, avoids human-induced errors, and allows agil reactions to changed business needs.
Often, it is considered as part of the requirement engineering process~\cite{dourdas2006discovering,mehandjiev2012cooperative,zachos2007discovering}, e.g., using ontologies~\cite{oriol2018ontology} or formal models~\cite{wang2012formal}.
Domains include Smart Cities~\cite{bensassi2018service} or cloud computing~\cite{jerbin2023request}.
Recent implementations also support OpenAPI specifications~\cite{apostolakis2023simple}.
Our work contributes to this knowledge corpus by analyzing how \glspl{llm} can be applied to the problem.

\subsection{\glsfmtshortpl{llm} for Automated Service Composition}
\glspl{llm} achieve remarkable results in natural language understanding, processing, and generation.
Initial ideas adopt an encoder-decoder architecture~\cite{radford2019better}.
Newer approaches use a decoder-only approach for text generation~\cite{radford2018improving} and encoder-only (embedding) models for similarity computation~\cite{devlin2019bert}.

Within \gls{soc}, we proposed initial concepts of using \glspl{llm} for service composition.
These still face the issues of input token limitations, imperfect results, and hallucinations~\cite{pesl2024verfahren,pesl2024uncovering,pesl2024compositio}.

Another approach to integrating \glspl{llm} with services (tools) is \gls{llm} agents.
These incorporate tool invocations into the chat interaction~\cite{mialon2023augmented,openai2024function,yao2023react}.
While facing similar problems like tool/service selection, the main difference to service composition is that the result of service composition is an executable, reusable artifact, e.g., as code.
In contrast, \gls{llm} agents invoke the tools directly during the answer creation~\cite{pesl2024compositio}.

A significant problem is the lack of appropriate benchmarks.
Although some initial proposals like RestBench~\cite{song2023restgpt} exist, a general benchmark across numerous domains is still missing.
We add to this by introducing generalized benchmarks.

\section{Research Method} \label{sec:research_methodology}

\begin{figure}
    \centering
    \includegraphics[width=\linewidth]{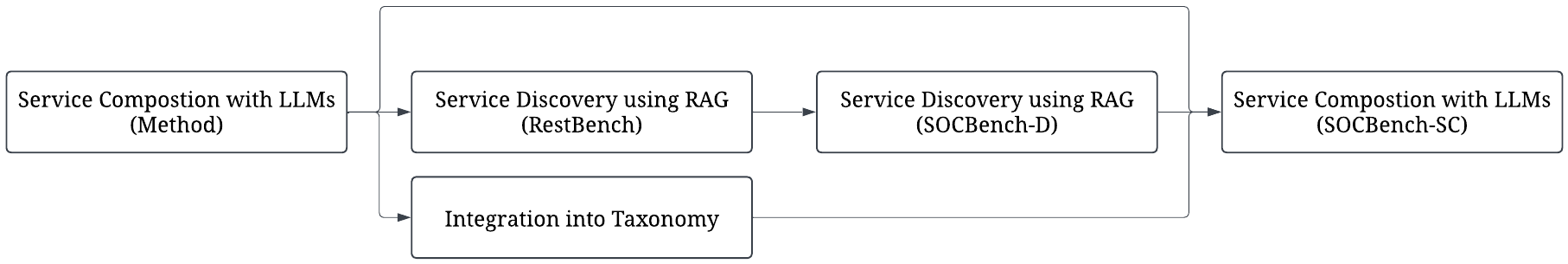}
    \caption{Research Methodology}
    \label{fig:research_methodology}
\end{figure}

\Cref{fig:research_methodology} shows our research methodology from left to right.
The first item is a general software architecture, i.e., a method to employ \glspl{llm} for automated service compositions.
Using this architecture, we can implement a prototype and measure the performance and implications of the approach.

Next, we look into the literature and analyze how to classify our method in the existing taxonomy of Lemos, Daniel, and Benatallah~\cite{lemos2015web} and how the taxonomy needs to be extended.
In parallel, we examine service documentation chunking, i.e., extracting the most relevant parts to allow using \gls{rag} for service discovery.
This allows us to mitigate prompt input token limitations.
We evaluate the \gls{rag} chunking approaches first by employing the exiting RestBench benchmark~\cite{song2023restgpt}, then by introducing our custom SOCBench-D benchmark generalized across all domains of the \gls{gics}~\cite{msci2023gics}.

Finally, we bridge the gap between service discovery and our software architecture by creating a benchmark comprising services based on the \gls{gics} domains, measuring the end-to-end performance from prompt to final service composition.
Further experiments could contain user studies to measure the time-saving of employing \glspl{llm} versus manual labor or applying \gls{llm} agents to improve reasoning.

\section{Contributions} \label{sec:contributions}
Following our research methodology from \Cref{fig:research_methodology}, we introduce four contributions.
The contributions are as follows:
\begin{enumerate}[noitemsep,topsep=0pt]
    \item The Compositio Prompto software architecture.
    \item Our extension to Lemos' taxonomy, which results in an extended taxonomy.
    \item The service discovery using \gls{rag}.
        This comprises the analysis of \gls{rag} using the existing RestBench~\cite{song2023restgpt} and the creation of the SOCBench-D benchmark.
        The resulting artifacts are a query-based service discovery benchmark and an algorithm that can dynamically create such a benchmark.
    \item The SOCBench-SC service composition benchmark.
        It contains the analysis of current \glspl{llm}, the benchmark itself, and the benchmark creation algorithm.
\end{enumerate}

\section{Preliminary Results} \label{sec:preliminary_results}

We already worked on the method, the taxonomy, and the service discovery with \gls{rag}.
Open points are in the full service composition and subsequent studies.

\subsection{Compositio Prompto}
\begin{figure}[h]
    \centering
    \includegraphics[width=0.85\linewidth]{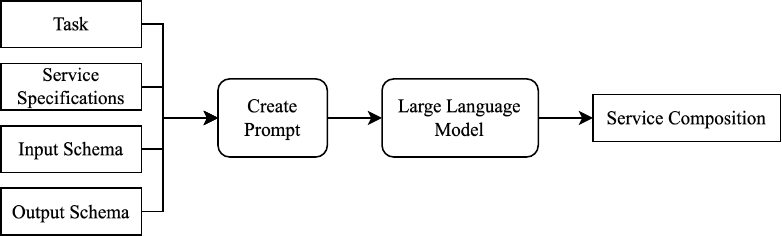}
    \caption{Compositio Prompto Architecture~\cite{pesl2024compositio}}
    \label{fig:compositio_prompto}
\end{figure}

First, we introduce our architecture \enquote{Compositio Prompto} to employ \glspl{llm} for automated service composition shown in \Cref{fig:compositio_prompto}~\cite{pesl2024compositio}.
It uses a task, service documentation, and an in- and output schema as input to create a prompt.
The prompt is then fed into the \gls{llm}, creating an executable service composition.

To evaluate Compositio Prompto, we implemented a fully operation prototype and performed a case study from the automotive domain.
Our prototype uses a natural language query as a task, OpenAPI specifications as service documentation, JSON schemas as in- and output schemas, and Python code for the executable service composition.
The results show that currently, only large models with more than 70B parameters can solve at least some tasks perfectly.
Nevertheless, even small models like Llama 3 8B produce close approximations of our manually crated sample solution.
Therefore, we conclude that current \glspl{llm} can be used to create a code recommendation that only needs little adaption, i.e., working time, to become operational.
Further research is needed to achieve full automation.

\subsection{Extended Taxonomy}
To analyze whether service composition with \glspl{llm} can already be expressed using the well-known taxonomy of Lemos, Daniel, and Benatallah~\cite{lemos2015web}, we examine for each category if it already covers the \gls{llm} capabilities adequately.
The result is an extended taxonomy, which comprises the additional subcategories needed to include the \gls{llm} capabilities in natural language understanding, semantic processing, and text generation~\cite{pesl2025taxonomy}~(submitted).

We validated our taxonomy extension using four existing \gls{llm}-based approaches for service composition from literature.
They show that each introduced new subcategory is indeed necessary.
Limitations are primarily in the original methodology, which only includes taxonomy elements that stem from actual service composition approaches.
Further extensions may be necessary once novel service composition approaches arise.

\subsection{OpenAPI \glsfmtshort{rag}}
The Compositio Prompto experiments highlight the challenge of the limited input token length of \glspl{llm}.
It leads to only being able to input only parts of the service documentation.
An example is that the OpenAPI of Spotify alone does not fit into the context size of OpenAI's GPT4o, leaving alone inputting additional services.
A technique to mitigate this issue is \gls{rag}, which splits the input data into smaller chunks.
The \gls{rag} system performs a semantic search using an embedding model based on these chunks and inserts the most relevant chunks into the prompt.
The benefit is that the inserted chunks are much smaller than the complete input data while revealing only the most relevant information~\cite{lewis2020retrieval}.

We apply \gls{rag} to service discovery to determine the influence of the model and chunking strategy~\cite{pesl2025rag}~(to appear).
We rely on the RestBench benchmark~\cite{song2023restgpt}, which consists of the Spotify and TMDB OpenAPI specification and pairs of natural queries with expected endpoints.
Our results show that it is beneficial to split the OpenAPI by endpoints~\cite{pesl2025rag}.

Nonetheless, the lack of general benchmarks arises.
Therefore, we extended our evaluation further by introducing the novel service discovery benchmark SOCBench-D, which generalizes across the \gls{gics} domains~\cite{pesl2025ragtsc}~(submitted).
We execute it using OpenAI's \textit{text-embedding-3-large}, Nvidia's \textit{NV-Embed-v2}, and BGE's \textit{bge-small-en-v1.5}.
Our results show that the choice of the chunking strategy is insignificant across all domains.
The number of received chunks is particularly relevant to the overall performance.
The second factor is the embedding model.
The Nvidia model outperforms the OpenAI model, which outperforms the BGE model.
Nonetheless, the BGE model reveals reasonable results.
In practice, this leads to the consideration that when resources are sparse, the BGE model can be used; when familiar with the OpenAI tooling, the OpenAI model can be used; and when interested in the best results, the Nvidia model can be employed.

\section{Future Work} \label{sec:future_work}
Our next research effort will comprise the analysis of complete service compositions incorporating the \gls{rag}-based service discovery.
The basic idea is to use a query-based service discovery benchmark like RestBench or SOCBench-D and extend it to code analysis.
This can be done by static code analysis, unit test-like testing, or creating custom mock instances of the services and tracking invocations.
The optimal approach has still to be determined.
The result is a benchmark for natural language service composition approaches, generalized across the \gls{gics} domains.

Further, we want to measure actual implications on development time savings, sustainability aspects, and advanced approaches like \gls{llm} agents.
These allow logical reasoning, which may reduce hallucinations and improve the composition quality.

\section{Concluding Remarks} \label{sec:conclusion}
With this Ph.D. thesis, we want to analyze the potential of employing \glspl{llm} for the well-known yet unresolved problem of automated service composition.
We introduced the Compostio Prompto architecture to realize automated service composition with \glspl{llm} in practice.
Further, we examined the usage of \gls{rag} for service discovery first by using the real-world RestBench benchmark and then by our cross-domain SOCBench-D benchmark.
Next, we will create a benchmark for general service composition cases and analyze how well current \glspl{llm} perform.
Other open points are the improvement of the result generation, e.g., by employing \gls{llm} agents, and the analysis of applicability in practice, e.g., by performing a user study.

\begin{credits}
\subsubsection{\ackname}
I want to thank my supervisor, Prof. Dr. Marco Aiello, for his ongoing support throughout my Ph.D. journey.
This preprint has not undergone peer review (when applicable) or any post-submission improvements or corrections. The Version of Record of this contribution is published in Intelligent Information Systems. CAiSE 2025. Lecture Notes in Business Information Processing, vol 557. Springer, Cham., and is available online at \url{https://doi.org/10.1007/978-3-031-94590-8_37}.

\subsubsection{\discintname}
The author is listed as inventors of a patent~\cite{pesl2024verfahren}, which covers \thetool for the automotive domain.
\end{credits}


\begin{thebibliography}{10}
\providecommand{\url}[1]{\texttt{#1}}
\providecommand{\urlprefix}{URL }
\providecommand{\doi}[1]{https://doi.org/#1}

\bibitem{apostolakis2023simple}
Apostolakis, I., Mainas, N., Petrakis, E.G.: Simple querying service for {OpenAPI} descriptions with semantic extensions. Information Systems  \textbf{117},  102241 (2023)

\bibitem{bensassi2018service}
Ben-Sassi, N., et~al.: Service discovery and composition in smart cities. In: Information Systems in the Big Data Era. pp. 39--48. Springer (2018)

\bibitem{berardi03}
Berardi, D., et~al.: {Automatic Composition of E-services That Export Their Behavior}. In: {ICSOC}. pp. 43--58 (2003)

\bibitem{bianchini2006ontology}
Bianchini, D., et~al.: Ontology-based methodology for e-service discovery. Information Systems  \textbf{31}(4),  361--380 (2006)

\bibitem{curbera2002unraveling}
Curbera, F., et~al.: Unraveling the web services web: an introduction to {SOAP}, {WSDL}, and {UDDI}. IEEE Internet Computing  \textbf{6}(2),  86--93 (2002)

\bibitem{degiacomo13}
{De Giacomo}, G., Patrizi, F., Sardi{\~n}a, S.: Automatic behavior composition synthesis. Artificial Intelligence  \textbf{196},  106--142 (2013)

\bibitem{devlin2019bert}
Devlin, J., Chang, M.W., Lee, K., Toutanova, K.: {BERT}: Pre-training of deep bidirectional transformers for language understanding. In: NAACL-HLT 2019. pp. 4171--4186 (2019)

\bibitem{dourdas2006discovering}
Dourdas, N., et~al.: Discovering remote software services that satisfy requirements: Patterns for query reformulation. In: CAiSE. pp. 239--254. Springer Berlin Heidelberg, Berlin, Heidelberg (2006). \doi{10.1007/11767138_17}

\bibitem{jerbin2023request}
Jerbi, I., et~al.: Request relaxation based-on provider constraints for a capability-based naas services discovery. In: CAiSE. pp. 611--627. Springer (2023)

\bibitem{lemos2015web}
Lemos, A.L., Daniel, F., Benatallah, B.: Web service composition: A survey of techniques and tools. ACM Comput. Surv.  \textbf{48}(3) (dec 2015)

\bibitem{lewis2020retrieval}
Lewis, P., et~al.: Retrieval-augmented generation for knowledge-intensive {NLP} tasks. In: NeurIPS. vol.~33, pp. 9459--9474. Curran Associates (2020)

\bibitem{mcDermott02}
McDermott, D.V.: {Estimated-Regression Planning for Interactions with Web Services}. In: AIPS. pp. 204--211. AAAI (2002)

\bibitem{mehandjiev2012cooperative}
Mehandjiev, N., L{\'e}cu{\'e}, F., Carpenter, M., Rabhi, F.A.: Cooperative service composition. In: CAiSE. pp. 111--126. Springer (2012)

\bibitem{mialon2023augmented}
Mialon, G., et~al.: Augmented language models: a survey (2023), \url{https://arxiv.org/abs/2302.07842}

\bibitem{msci2023gics}
{MSCI Inc.}, {Standard \& Poor's}: Global industry classification standard ({GICS}). \url{https://www.msci.com/gics} (August 2024)

\bibitem{openai2024function}
OpenAI: Function calling and other {API} updates (Jun 2024), \url{https://openai.com/index/function-calling-and-other-api-updates/}, last accessed 2024-07-18

\bibitem{openapi2021spec}
{OpenAPI Initiative}: {OpenAPI} specification. \url{https://www.openapis.org/} (2021), version 3.1.0, last accessed 2025-03-15

\bibitem{oriol2018ontology}
Oriol, X., Teniente, E.: An ontology-based framework for describing discoverable data services. In: CAiSE. pp. 220--235. Springer, Cham (2018)

\bibitem{pesl2025taxonomy}
Pesl, R.D., Aiello, M.: Revisiting {Lemos'} taxonomy for service compositions with large language models. In: ICWS (2025), submitted

\bibitem{pesl2024verfahren}
Pesl, R.D., Klein, K., Aiello, M.: {Verfahren} zur {Nutzung} von unbekannten neuen {Systemdiensten} in einer {Fahrzeuganwendung} (2024), {Patent DE102024108126A1}

\bibitem{pesl2025rag}
Pesl, R.D., Mathew, J.G., Mecella, M., Aiello, M.: Advanced system integration: Analyzing {OpenAPI} chunking for retrieval-augmented generation. In: CAiSE (2025), \url{https://arxiv.org/abs/2411.19804}, to appear

\bibitem{pesl2025ragtsc}
Pesl, R.D., Mathew, J.G., Mecella, M., Aiello, M.: Retrieval-augmented generation for service discovery: Chunking strategies and benchmarking. TSC  (2025), submitted

\bibitem{pesl2024uncovering}
Pesl, R.D., Stötzner, M., Georgievski, I., Aiello, M.: Uncovering {LLMs} for service-composition: Challenges and opportunities. In: ICSOC 2023 WS. Springer (2024)

\bibitem{pesl2024compositio}
Pesl, R.D., et~al.: {Compositio} {Prompto}: An architecture to employ large language models in automated service computing. In: ICSOC 2024. pp. 276--286. Springer (2025)

\bibitem{radford2019better}
Radford, A., Wu, J., Amodei, D., Amodei, D., Clark, J., Brundage, M., Sutskever, I.: Better language models and their implications. OpenAI blog  \textbf{1}(2) (2019), \url{https://openai.com/index/better-language-models/}, last accessed 2024-11-28

\bibitem{radford2018improving}
Radford, A., et~al.: Improving language understanding by generative pre-training (2018)

\bibitem{shesh03}
Sheshagiri, M., DesJardins, M., Finin, T.: {A planner for composing services described in DAML-S}. In: 13th ICAPS WS on planning for Web services (2003)

\bibitem{10.1007/978-3-031-57853-3_3}
Soki, A.T., Siqueira, F.: Discovery of {RESTful} {Web} services based on the {OpenAPI} 3.0 standard with semantic annotations. In: AINA. pp. 22--34. Springer (2024)

\bibitem{song2023restgpt}
Song, Y., et~al.: {RestGPT}: Connecting large language models with real-world {RESTful} {APIs} (2023), \url{https://arxiv.org/abs/2306.06624}

\bibitem{wang2012formal}
Wang, H.H., et~al.: A formal model of the semantic web service ontology {(WSMO)}. Information Systems  \textbf{37}(1),  33--60 (2012)

\bibitem{yao2023react}
Yao, S., et~al.: React: Synergizing reasoning and acting in language models (2023), \url{https://arxiv.org/abs/2210.03629}

\bibitem{zachos2007discovering}
Zachos, K., et~al.: Discovering web services to specify more complete system requirements. In: CAiSE. pp. 142--157. Springer (2007)

\end{thebibliography}
\end{document}